\documentclass[fleqn,twoside]{article}
\usepackage[headings]{espcrc2}

\readRCS
$Id: espcrc2.tex,v 1.2 2004/02/24 11:22:11 spepping Exp $
\usepackage{graphicx}
\usepackage[figuresright]{rotating}

\newcommand{\AmS}{{\protect\the\textfont2
  A\kern-.1667em\lower.5ex\hbox{M}\kern-.125emS}}

\title{The hamiltonian study of supersymmetric Yang-Mills quantum mechanics}

\author{M. Trzetrzelewski\address{M. Smoluchowski  Institute of Physics, Jagiellonian University, \\
Reymonta 4, 30-059 Cracow, Poland \\ }%
        \thanks{I thank the organizers of the Cargèse Summer School, May 22-June 3, 2006 for their kind invitation and support.}
     }

\runauthor{M. Trzetrzelewski}

\begin{document}

\begin{abstract}
The hamiltonian  formulation of Supersymmetric Yang-Mills quantum
mechanics (SYMQM) is discussed. We focus on the Fock space
formulation of the models since it is convenient for the numerical
analysis, however some novel analytical results are also pointed
out.

\vspace{1pc}
\end{abstract}

% typeset front matter (including abstract)
\maketitle

\section{A brief history of SYMQM}

Supersymmetric Yang-Mills quantum mechanics by  definition are
$\mathcal{N}=1$ supersymmetric Yang-Mills field theories reduced
from $D=d+1$ spacetime dimensions to $0+1$ dimensions.

Almost thirty years ago the purely bosonic part of SYMQM was
conceived by Bjorken  \cite{Bjorken} in the zero volume limit of
YM theories. The subject was then pushed forward by L\"usher
\cite{Lusher} resulting in small volume expansion of glueball
masses for the $SU(2)$ gauge group. At the same time the model was
considered by Hoppe in his Phd. thesis as the regularized
description of relativistic membrane \cite{Hoppe}. The
supersymmetric formulation of YMQM  was first discussed by
Cloudson and Halpern \cite{Claudson}. Later on the supermembrane was formulated \cite{Polchinski,Sezgin}
as well as its regularized description \cite{deWitt}
resulting precisely in SYMQM. The interest in quantum formulation
of (super)membranes was originally motivated by particle physics
i.e. membranes were believed to describe elementary particles, the
idea first put forward by Dirac \cite{Dirac}. However, it turns
out that the spectrum of the hamiltonian of SYMQM is continuous
hence the supermembrane is unstable \cite{Nicolai}. Almost a
decade later this set back was interpreted as good news since this
time SYMQM were conjectured \cite{BFSS} to describe M-theory. The
continuous spectrum corresponds now to the scattering states on
M-theory side and there is no contradiction since M-theory is a
second quantized theory.
\section{The Fock space approach}

Supersymmetry imposes some constraints on  dimensionality of SYM
theories ( hence SYMQM )  namely $D=2,3,4,6,10$. SYMQM consists of
real bosonic variables $x^i_a$ and  complex fermionic variables
$\psi^{\alpha}_a$ ( for $D=10$ the 16 Majorana-Weyl fermions can
be composed into the 8 complex ones ). Here $a=1,\ldots, N^2-1$ is
a color index i.e $x^i_a$ and $\psi^{\alpha}_a$ are in the adjoint
representation of $SU(N)$ while $i=1,\ldots,D-1$ and
$\alpha=1,\ldots,D-2$ are spatial and spinor indices respectively.
It seems that there is a mismatch between bosonic and fermionic
degrees of freedom however, while performing the dimensional
reduction the Gauss law becomes a global $SU(N)$ invariance. In
the subspace of $SU(N)$ singlets the degrees of freedom match. The
hamiltonian of SYMQM is then  \cite{Claudson}
\begin{equation}
H=\frac{1}{2}\pi^i_a\pi^i_a+\frac{1}{4}g^2(f_{abc}x_b^ix_c^j)^2+H_F,
\end{equation}
where $f_{abc}$ are $SU(N)$ structure constants, $\pi^i_a$ are
conjugate to $x^i_a$, $[x^i_a,\pi^j_b]=i\delta_{ab}\delta^{ij}$
and $H_F$ is the fermionic part which schematically is
$H_F=igx\bar{\psi}\Gamma \psi$, where $\Gamma$ are Dirac gamma
matrices in corresponding dimension.

We now introduce, in the standard fashion, the bosonic creation
and annihilation operators ${a^{\dagger}}^i_a$, ${a}^i_a$ and
analogously fermionic operators ${f^{\dagger}}^{\alpha}_a$,
$f^{\alpha}_a$ although in this last case the construction depends
on the dimensionality ( see \cite{D4,D10} for conventions ).  Any state $\mid s \rangle$ in Fock space
is now a linear combination of products of creation operators
acting on the Fock vacuum $\mid 0 \rangle $, so that the resulting
state $\mid s \rangle$ is gauge invariant. For numerical analysis
we now introduce a cutoff $n_{B_{max}}$ and truncate the Hilbert
space so that the states have no more then $n_{B_{max}}$ number of
bosonic quanta. Then we compute the matrix elements of (1) and
diagonalize the truncated matrix $H^{(n_{B_{max}})}$. The
resulting eigenvalues converge very fast to the exact eigenvalues
of the hamiltonian provided the spectrum is discrete. In the case
of continuous spectrum the situation is more subtle. We
refer to \cite{Wosiek33} where it is discussed in details.

The method just described was applied to $D=4$, $SU(2)$ SYMQM
resulting in very precise evaluation of the spectrum and the
Witten index \cite{D4}.  The analogous calculation for $D=10$
SYMQM is difficult to perform since in this case the number of
states from Fock space grows extremely fast with $n_{B_{max}}$
\cite{D10}. The remedy at this point could be the additional
$SO(9)$ invariance. It is then possible that the numerical
analysis of these models is within reach once we work in sectors
with given $SO(9)$ angular momentum.

\section{Exact results and large N}

While for SYMQM with $D>2$ there are no exact solutions the $D=2$
case is different. The system is much simpler then the higher
dimensional relatives since it is where the quartic potential term
in the hamiltonian, vanishes. The hamiltonian is simply
$H=\frac{1}{2}\pi_a\pi_a$ however, the model is not free due to the
Gauss law. The exact solutions in the bosonic sector are known for
$SU(2)$ \cite{Claudson} and arbitrary $SU(N)$ \cite{Samuel}
groups.
 The disadvantage of these
solutions is the absence of explicit $N$ dependence hence the $N
\to \infty$ limit is difficult.  Moreover, the existing solutions
are not general ones except for the $D=2$, $SU(2)$ SYMQM. It
is possible to overcome these difficulties. In \cite{33}
 we have found a class of solutions which differ
from the existing ones. Moreover, their large $N$ limit is possible although
 it turns out that one has to be
very careful in performing the limit due to the distinguished role
of the bilinear operators $x_ax_a$. The result reads
\begin{equation}
\mid p \rangle =  e^{-p^2r^2/2N^2}\mid v \rangle, \ \ \ \ r=\sqrt{x_ax_a}, \ \ \ \ N>>1,
\end{equation}
where $\mid v \rangle$ is the vacuum state and $p$ is the momentum. We see that all the
$1/N^k$ terms are present in the Taylor expansion of the large $N$
solutions. It can be argued that they are all of equal importance
hence the "ordinary" large $N$ techniques may fail when the
hamiltonian considered has continuous spectrum. However, this is
precisely the case for all SYMQM. Moreover, if we put $g=0$ in (1)
we obtain the class of large $N$ solutions, having the form of
products of (2), for SYMQM in arbitrary dimensions.

\section{Conclusions}

The numerical approach to SYMQM presented here is very
encouraging. Although so far applied only to $D=4$ case we believe
that it will finally give us the nonperturbative results also in the
$D=10$ system.

A careful large $N$ analysis  indicates that the $1/N^k$ terms may
play an important role which is in contrast with the majority of
large $N$ computations.

 \section{Acknowledgments}
This work was supported by the the grant of Polish Ministry of
Science and Education no. P03B 024 27 ( 2004 - 2007 ) and N202 044
31/2444 ( 2006-2007 ) and the  Jagiellonian University Estreicher
foundation.

\end{document}